\documentclass[prd,twocolumn,showpacs,superscriptaddress,nofootinbib,floatfix]{revtex4}

\usepackage[latin1]{inputenc}
\usepackage{amsmath,graphicx}

\begin{document}

\title{Self-organized criticality in boson clouds around black holes}

\author{Gabriela Mocanu}
\email{gabriela.mocanu@ubbcluj.ro}

\affiliation{Faculty of Physics, Department of Theoretical and
Computational Physics, Babes-Bolyai University, No. 1
Kog\u{a}lniceanu Street, 400084, Cluj-Napoca, Romania}
\affiliation{Institute for Theoretical Physics, Vienna University
of Technology, Wiedner Hauptstrasse 8-10/136, A-1040 Vienna,
Austria, Europe}

\author{Daniel Grumiller} \email{grumil@hep.itp.tuwien.ac.at}
\affiliation{Institute for Theoretical Physics, Vienna University
of Technology, Wiedner Hauptstrasse 8-10/136, A-1040 Vienna,
Austria, Europe}

\date{\today}

\preprint{TUW-12-xx}

\begin{abstract}
Boson clouds around black holes exhibit interesting physical
phenomena through the Penrose process of superradiance, leading to
black hole spin-down. Axionic clouds are of particular interest,
since the axion Compton wavelength could be comparable to the
Schwarzschild radius, leading to the formation of ``gravitational
atoms'' with a black hole nucleus. These clouds collapse under
certain conditions, leading to a ``Bosenova''. We model the
dynamics of such unstable boson clouds by a simple cellular
automaton and show that it exhibits self-organized criticality.
Our results suggest that the evolution through the black hole
Regge plane is due to self-organized criticality.
\end{abstract}


\maketitle


\section{Introduction}

Modern particle detectors are typically associated with very high
energies, like the current design of the LHC with 8 TeV center of
mass energy. The need for having high energies is obvious if the
particles to be discovered --- like the Higgs --- have a
substantial rest mass 
or, conversely, a tiny Compton wavelength.

Interestingly, for (QCD-)axions
\cite{Peccei:1977hh,Weinberg:1977ma,Wilczek:1977pj} the situation
may be reversed: the rest mass of axions can be tiny, while their
Compton wavelength can be substantial. More precisely, the axion
Compton wavelength can be of the same order of magnitude as the
Schwarzschild radius of a stellar mass black hole. Thus, at least
in principle, black holes can serve as particle detectors for
axions \cite{Arvanitaki:2009fg,Arvanitaki:2010sy}. This idea has
engendered a lot of recent interest
\cite{Acharya:2010zx,Marsh:2010wq,Panda:2010uq,Dubovsky:2010je,Marsh:2011gr,Dubovsky:2011tu,Kodama:2011zc,Cardoso:2011xi,Marsh:2011bf,Horbatsch:2011ye,Alsing:2011er,AmaroSeoane:2012km}.

The present work is based on the working assumption that suitable
axions exist in our Universe (we shall be more precise what
``suitable'' means in the body of the paper). In that case an
axion cloud forms around a rotating black hole, and the combined
system black hole/axion cloud can be thought of as a gigantic
``gravitational atom'', with the black hole playing the role of
the nucleus and the axion cloud playing the role of the electrons.

As explained in \cite{Arvanitaki:2010sy}, the dynamics of this
``gravitational atom'' is governed by the Penrose process of
superradiance \cite{Penrose:1969pc}. This process leads to
exponential growth of the occupation numbers of the ``atomic''
levels, i.e., the formation of an axion Bose--Einstein condensate.
The energy required for the formation of this condensate is
extracted from the black hole, which thereby spins down and also
loses some of its mass $M$. In fact, the black hole parameters
$(M,a)$ move on a Regge trajectory, where $a\in[0,1]$ is the Kerr
parameter. Eventually, when the attractive axion self-interactions
become stronger than the gravitational binding energy, the axion
cloud collapses --- a ``Bosenova'' occurs.

The aim of our present work is to map relevant aspects of the
dynamics of the black hole/axion cloud system to a simple cellular
automaton, and to show that it exhibits self-organized
criticality. If this conclusion holds also for more realistic
cellular automata it would mean that self-organized criticality
governs an essential part of the dynamics of gravitational atoms.

This paper is organized as follows. In section \ref{sec:2} we
review some preliminaries, namely crucial properties of
self-organized criticality and the black hole/axion cloud system.
In section \ref{sec:3} we present the working assumptions of
relevance for our cellular automaton and argue to what extent they
are plausible. In section \ref{sec:4} we present results based
upon computer simulations implementing this specific cellular
automaton. In section \ref{sec:5} we summarize and point to open
issues.

\section{Preliminaries}\label{sec:2}

\subsection{Self-organized criticality}

The concept of Self Organized Criticality (SOC) was proposed as an
organizing principle of Nature, as it was realized that numerous
spatial extended systems, in physics, biology, social media and
economics, exhibit a number of properties which may be shortly
characterized as flicker noise for the temporal evolution and
self-similar (fractal) behavior for the spatial evolution
\cite{Bak:1988zz}. Ever since this seminal paper, the analytical
and numerical methods developed to be able to tackle SOC in
various contexts have led to progress in understanding the
nonlinear dynamics of complex, interacting systems. It is expected
that SOC develops in slowly driven interaction dominated threshold
systems and it has been shown that this state is an attractor of
the dynamics of such systems (see for instance \cite{Pruessner}
and references therein).

Due to the complex interactions between its constituents, after a
transitory phase in its evolution, the system exhibits what may be
called a phase transition to a state characterized by the absence
of a critical length scale, i.e., in this state all lengthscales
are relevant.

The archetypal SOC system is driven by an external force with a
very large characteristic timescale and the system evolution is
determined by local rules, i.e., only near-neighbor interactions
are considered \cite{Sornette,Pruessner}. One important point to
make is that this external driver is not artificial/conscious but
is intrinsic to the physics in the system. This property of SOC is
unlike the phase transitions in an Ising type system, where the
phase transition is acquired by a conscious tuning of parameters
to a critical value.

Because this line of research is still in its infancy and due to
the associated analytical complexity, analysis of such systems
relies heavily on numerical methods and so they have become
intertwined in an inseparable way. It is difficult to enumerate
properties of SOC without relating them to the way they are
simulated in computer experiments.

Systems which exhibit SOC are \emph{discrete} (in general modelled
by lattice models). The \emph{interaction between constituents}
are usually nearest neighbor-like and governed by local rules and
are modelled as interaction between sites in a lattice, such that
if one site receives an information, only a few other selected
sites have access to this information. The entire system is
subjected to an infinitely slow external \emph{drive} and due to
this drive and the complex interaction, the onset of SOC usually
occurs when all the sites in the lattice host a
near-\emph{threshold} value of some parameter. If the threshold is
reached, the system becomes subjected to different rules of
evolution (critical flow) which allow information to be
transmitted not only locally, but across all lengthscales. If one
site has reached threshold and sends information to the next site,
this next site might also reach threshold and so on. The
avalanches of information from one seed site to a final site (i.e.
where the threshold condition is no longer satisfied) form an
event. The size of an event is the total number of avalanches
occurring from one seed site. Natural systems that reach SOC
maintain it for an indefinite time as long as the external driver
keeps acting. This equilibrium state is maintained by a balance
between driving and \emph{dissipation}, which occurs at the
boundaries or in the bulk.

The way that SOC was historically ``detected" in Nature was due to
the properties exhibited by observables of the system. These
observables show a \emph{powerlaw} distribution of values. It was
shown~\cite{Sornette,Pruessner} that, for a system in SOC, the
distribution function $D(s)$ of some parameter $s$ is given by

\begin{equation}
D(s) = a s^{-\tau}\mathcal{G}\left(\frac{s}{bL^\sigma}\right),\label{eq:distrib}
\end{equation}
where $\mathcal{G}(x)$ is the universal scaling function for a
given universality class of systems, $a$ is a system specific
constant, $\tau$ is the critical exponent, $b$ is a system
dependent amplitude, $L$ is the system spatial size and $\sigma$
is determined by the physical dimension of the parameter $s$. A
very useful feature of SOC is that if the distribution of another
observable, say $y$, is considered, then $D(y)$ will be equivalent
to Eq.~\eqref{eq:distrib}, with $s$ replaced by $y$ and different
parameters $a$, $\tau$, $b$ and $\sigma$, but the \textbf{same}
scaling function $\mathcal{G}$.

This is conceptually a powerful tool for observational purposes,
in cases where it proves easy to record $D(s)$, but difficult to
record a perhaps more interesting distribution $D(y)$.

An example of a system in SOC and a discussion of the properties
emphasized above is given in the Appendix.

\subsection{Massive boson instabilities near a rotating black hole}

It has been
shown~\cite{Arvanitaki:2010sy,Kodama:2011zc,Zouros:1979iw} that
the presence of bosons near a rotating black hole (BH) will lead
to the formation of a boson cloud (BC) around the BH if a series
of conditions are met. This BC-BH system is analogous to an atom.
The cloud is being continuously fed with bosons due to the
superradiance phenomenon. At some point, when the number of bosons
in the cloud is too large, the nonlinearities in the BC lead to an
unstable behavior and the cloud goes through a Bosenova-type
event.

The conditions for this chain of events to occur refer to a set of
parameters characterizing both the boson and the BH, $\{ M, \omega
_h, \mu, \omega _R, m \}$, where $M$ and $\omega _h$ are the mass
of the BH and the angular velocity at the horizon, respectively,
and $\mu$, $\omega _R$ and $m$ are the mass of the boson, angular
velocity of the associated wavepacket and ``magnetic'' quantum
number of the boson. The first condition requires that a boson
exists such that its associated Compton wavelength, $\lambda _b$,
is approximately equal to the characteristic size of the BH, $R_g$
(the Schwarzschild radius). It was shown
that the QCD axion can satisfy this condition~[in e.g. \cite{Arvanitaki:2010sy}]. Superradiance,
i.e., continuous feeding of the cloud with ``free'' bosons, the
phenomenon analogous to the Penrose process for fields, occurs if
the superradiance condition is met, namely if $\omega \in [0,
\omega _h m]$. Once superradiance sets in, the topology of the
growing cloud is given by a frequency $\omega _R$, which satisfies
the superradiance condition and is the frequency of
hydrogenic-like bound levels. Growth of the BC, instability and
subsequent Bosenova occur if the frequency associated to the boson
\emph{as it interacts with the BH} has a very small imaginary
part, $\Gamma _g$, such that the frequency describing the
wavepacket in the potential well of the BH is $\omega = \omega _R
+ \imath \Gamma _g$. The parameter $\Gamma _g$, giving the growth
of the BC, is a function of the set of parameters mentioned above
and it exhibits a maximum for specific values of these
parameters~\cite{Arvanitaki:2010sy}.

We are now in the position to define the ``suitable axion''
mentioned in the Introduction. This axion is a boson which
possesses physical characteristics such that $\lambda _{b} \simeq
R_g$, $\omega _R \in [0, \omega _h m]$ and, in the potential well
of a BH, exhibits a growth rate $\Gamma _g$ which is very small,
$\omega _R \gg \Gamma _g$. The growth rate also depends on details
of the bound system, such as the hydrogenic-like quantum numbers,
taken here such that $\Gamma _g$ has its maximum possible value. We assume that
this boson exists and give a general outline of the dynamics of
the BC-BH system. Superradiance sets in and the number $N$ of such
bosons in the vicinity of the BH is amplified, in the linear
limit, as

\begin{equation}
\frac{dN}{dt} = \Gamma _g N.\label{eq:lin}
\end{equation}

However, during this amplification more complex processes occur.
Most importantly, the nonlinearities start increasing and the
bosons begin to interact with each other. The end result is that
the number of axions present in the cloud is reduced, with a rate
$\Gamma _d$ ($d$ from dissipation), where $\Gamma _d \ll \Gamma
_g$. For our purposes, Eq.~\eqref{eq:lin} can be naively modified
as

\begin{equation}
\frac{dN}{dt} = \Gamma _g N - \Gamma _d N.\label{eq:linDiss}
\end{equation}

Even in this very simple description it is possible to recover
some of the qualitative results derived rigorously
in~\cite{Arvanitaki:2010sy}. Since dissipation does not
equilibrate growth, the number $N$ of bosons grows until the
back-reaction on the frequency $\omega _R$ is such that the shape
of the cloud departs from hydrogenic (nonlinear limit). This has
been shown~\cite{Arvanitaki:2010sy,Kodama:2011zc} to occur when

\begin{equation}
\epsilon \equiv \frac{M_{BC}}{M_{BH}}= 10^{-4},\label{eq:epsilon}
\end{equation}
where $M_{BC}$ is the mass of the boson cloud.

The cloud grows further, but in a nonlinear regime, until the
nonlinearities in the boson self interaction potential are too
large and the cloud is no longer stable. A large percentage of the
cloud is supposed to
collapse~\cite{Arvanitaki:2010sy,Kodama:2011zc} (critical limit).
This collapse occurs on a timescale of $r_g$ of the BH, meaning
that from the moment the critical limit has been reached, it takes
$r_g$ units of time until a significant part of the cloud
collapses into the BH.

The subtlety here is that this process occurs following an
avalanche of information spreading on all lengthscales accessible
to the system (i.e. with an upper bound given by the system size).

If the superradiance condition is still met, the cloud regrows
within the same parameter space $\{ M, \omega _h, \mu, \omega, m
\}$. In a plot of the BH spin vs. $M\mu$ (a Regge plot,
\cite{Arvanitaki:2010sy}, their Fig.~3), the subsequent collapses
and regrowths, with the same values of the parameters that
initially satisfied the superradiance condition, resemble an
almost straight line on which the systems wanders (Regge
trajectories). It stays on this line for tens to hundreds of
$e$-fold times of boson cloud growth, i.e. for tens to hundreds of
Bosenovas. When the superradiance condition is no longer met with
the initial parameters, the system makes a transition to another
set of parameters (i.e., a new line on the Regge plot) which are
again superradiant and analogous dynamics occurs.

\section{Cellular automaton for boson cloud around black holes}\label{sec:3}

We want to map the dynamics of BC-BH system into SOC. As was
mentioned previously, the concept of SOC and the simulation of SOC
are intimately connected and the only contained way to map the
BC-BH dynamics to SOC is to simultaneously discuss the computer
simulation associated with this analysis. SOC simulations are
generally done using Monte Carlo methods on discrete grids, and we
will focus on a Cellular Automata (CA) approach. The CA is a
``machinery'' which stores values of a parameter on a discrete
grid and evolves these values in time and space according to
simplistic rules designed to mimic a real
phenomenon~\cite{Bandman:2008}.

The CA used in the remainder of this paper is 2-dimensional
cartesian grid, of size $L_x \times L_y$. While this undoubtedly
simplifies the numerical effort, there is also strong physical
argumentation in favor of considering a 2D grid (instead of a 3D
grid). The BH has a significant spin, the quantum numbers of the
atom are fixed (by assumption of the ``suitable axion'') and the
orbits are Keplerian, which means that angular momentum is
conserved thus leading to planar motion.

Three questions then arise: 1) what is the quantity which will be
placed on the grid, 2) whether it is physically acceptable to
discretize that quantity in this manner and 3) what are the
evolution rules? We proceed by answering the first two questions
here, while the third one will be the subject of
Section~\ref{sec:4}.

Our simulation procedure is based on the following reasoning: if
the number of axions in the cloud can be computed by some means
[as in~\cite{Arvanitaki:2010sy}, their Eq.~(28)], a local quantity
can be defined by dividing the number of axions by the area of the
cloud. Because the area of the cloud is considered to be constant,
simulating the evolution of the density means simulating the
evolution of the number of axions, and this is why we will refer
to the quantity in the grid as ``the number of bosons''.

This basically means that the $N$ bosons in the $L \times L$ cloud
will be accommodated into a $\bar{L}\times \bar{L}$ grid, where we
will from now on denote $L_x=L_y=\bar{L}$. So it follows that one
cell in the grid has $\bar{N} = N/\bar{L}^2$ bosons and $\bar{N}$
will be the parameter stored in the CA.

\begin{figure}[!h]
\centerline{\includegraphics[width=\linewidth]{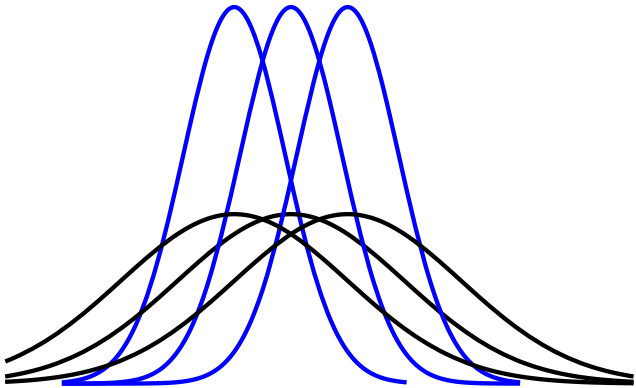}}
\caption{Schematic representation of the distribution of
information about the density in the cloud. Black lines: $N$
poorly localized bosons. Blue lines: $N$ "effective" bosons,
better localized for simulation purpose.}\label{fig:mental}
\end{figure}

To answer the more important question if this discretization is
physically acceptable, consider that the number of bosons, $N$, or
more precisely the $N$ very poorly localized wavefunctions allowed
by the superradiance, may be replaced by wavefunctions which peak
sharply, i.e. are more localized for simulation purposes
(Fig.~\ref{fig:mental}).  This ``effective'' system has the same
information content from the point of view of an outside viewer
looking at gravitational radiation following collapse. One will
see in the following algorithm/simulations that the information
released following collapse does not depend on the actual position
of where the information was released. Thus, from the point of
view of an external observer equipped only to record information
on gravitational radiation, it is not important if an axion left
the cloud by withdrawing its wavefunction from the entire cloud or
only from a more localized spatial extent as in our simulation.

We address now how the temporal evolution of the BC-BH system looks
like in this framework. The linear dynamics of the BC-BH are
implemented and are the evolution rules of the CA. $\bar{N}$ is
increased at randomly chosen positions at each timestep in order
to simulate superradiance and $\bar{N}_{i,j}(k)$ is the value of
the variable at the grid-cell labelled by spatial indices $i$ and
$j$, at time step $k$, as explained in detail in
Section~\ref{sec:4}. We record the value of a given parameter $s$
characterizing the system and check if the distribution function
of this parameter, $D(s)$, behaves as for a SOC system.

We summarize our assumptions

\begin{itemize}

\item{}There is only one populated energy level. This is
equivalent to saying that Eq.~\eqref{eq:linDiss} expresses, in a
linear approximation, the dynamics of the cloud itself. Such an
assumption is justified, because the growth rate of the next
available energy level is very small compared to $\Gamma _g$.

\item{}The dissipation rate $\Gamma _d$ is constant as the cloud
density grows. By assumption, dissipation is negligible compared
to growth, so even if $\Gamma _d$ does change, it will not change
as much as to affect the dynamics~\cite{Arvanitaki:2010sy}.

\item{}The nonlinear and critical limit are the same and occur
when the parameter $\epsilon$ reaches its critical value
$10^{-4}$, Eq.~\eqref{eq:epsilon}. This is supported by
simulations of collapse~\cite{Kodama:2011zc}.

\item{}the size of the cloud, $L$, is constant and is calculated
based on the linear approximation of hydrogenic states.

\end{itemize}

The mapping between BC-BH and SOC/CA proposed in this paper is
summarized in Table~\ref{table:map}.

\begin{table}\caption{Map of BC-BH dynamics to SOC.}
\begin{tabular}
{|p{0.5\linewidth}|p{0.5\linewidth}|}\hline
    SOC/CA framework & BC-BH system \\ \hline
    discrete space & 2D ``cloud'' \\ \hline
    local interaction & one cloud cell interacts with 3 other cloud
cells only at criticality \\ \hline
    infinitely slow external drive & bosons are added to the cloud
with a very slow rate, $\Gamma_g$ \\ \hline
    SOC occurs as a result of threshold dynamics & when the critical
parameter $\epsilon = 10^{-4}$ is reached, nonlinearities become
important \\ \hline
    dissipation & given by the very slow rate
$\Gamma_d$ \\ \hline
    observables & distribution of the number of
avalanches needed to relax one perturbation \\ \hline
\end{tabular}\label{table:map}\end{table}

\section{Results}\label{sec:4}

We address now the third question posed in the previous Section by
formulating the evolution rules for our CA. Let $L$ be the cloud
radius, i.e., the $r_c$ quantity from Eq.~(11) of
\cite{Arvanitaki:2010sy}

\begin{equation}
L \equiv r_c = \frac{(n+l+1)^2}{(\mu R_g) ^2}R_g,
\end{equation}
with $n=0$, $l=1$, $\mu R_g = 0.42$ and $\mu  = 3 \cdot
10^{-11}eV$. $n$ and $l$ are quantum numbers associated with the
hydrogenic state of the cloud and these particular values assure
maximum growth. In this case, the constant~\cite{Gerton:2000} most
effective instability rate occurs when \cite{Arvanitaki:2010sy}

\begin{equation}
\Gamma _g = 1.5 \cdot 10^{-7} rg^{-1}.
\end{equation}

The solution of the differential equation~\eqref{eq:lin} is given
by

\begin{equation}
N(t) = N_i e^{\Gamma _g t},
\end{equation}
where $N_i$ is the initial value of the variable $N$. For
simulation purposes, temporal evolution is recorded in discrete
timesteps, $\Delta t$. The solution at timestep $k$ may be
rewritten as

\begin{equation}
N(k \cdot \Delta t) = N_i e^{\Gamma _g k \Delta t}.
\end{equation}

If $\Delta t$ is taken to be of the order of one gravitational
radius and the notation $\bar{\Gamma} _g = \Gamma _g \Delta t$ is
used, the equation for the evolution of the system in natural time
units is formally written as

\begin{equation}
N(k) = N_i e^{\bar{\Gamma} _g k}.
\end{equation}

The number of bosons added to the cloud due
to superradiance in one natural timestep is given by

\begin{equation}
\Delta N = N(k+1)-N(k) = N_0 e^{\bar{\Gamma} _g k} \left
(e^{\bar{\Gamma} _g} -1 \right ).
\end{equation}

Since for the QCD axion $e^{k\bar{\Gamma} _g} \cong 1$ and
$e^{\bar{\Gamma} _g} -1 \cong 1.5 \cdot 10^{-7}$, the dynamics
equation used as a basis for the simulation is

\begin{equation}
N(k+1) = N(k) + 1.5 N_i \cdot 10^{-7}.
\end{equation}
It is useful to restate this equality in terms of the critical
number of bosons, $N_c$. We will do this by assuming that the
initial number of bosons is some fraction $x \in (0,1)$ of the
critical number, $N_i = x N_c$

\begin{equation}
N(k+1) = N(k) + 1.5 x N_c \cdot 10^{-7}.
\end{equation}

$N_c$ is easily computable as $N_c=M_{BH}\epsilon/\mu=2.4\times
10^{16}$ (all quantities are calculated for mass expressed in eV).
So the equation for the temporal evolution of the total number of
bosons in the cloud can be written as

\begin{equation}
N(k+1) = N(k) + \Delta N,
\end{equation}
where $\Delta N = 3.6 x \cdot 10^{9}$. This is to be simulated
on a $100 \times 100$ grid with the help of the parameter
$\bar{N}$. One may immediately write

\begin{equation}
\bar{N}_{i,j}(k+1) =\bar{ N}_{i,j}(k) + \Delta \bar{N},
\end{equation}
for \emph{all} sites labelled $i,j$ in the grid and at each
timestep $k$. This is not general enough because so far we have no
reason to believe that the spatial growth of the number of bosons
is homogeneous. To circumvent this problem, the same total amount
$ \Delta N$ will be added to the cloud, but to a limited number of
randomly chosen sites, i.e. the update rule for the CA will read

\begin{equation}
\bar{N}_{i,j}(k+1) =\bar{ N}_{i,j}(k) + \bar{N}_0,
\end{equation}
for $n$ randomly chosen sites, with $N_0=\Delta N/n$. If $n=100$
then $\bar{N}_0=3.6 x \cdot 10^7$. For the discretized case the
critical parameter is $\bar{N_c}=2.4 \cdot 10^{12}$. Dissipation
must also be taken into account, but at a much smaller rate. This
is modelled as

\begin{equation}
\bar{N}_{i,j}(k+1) =\bar{ N}_{i,j}(k) - \bar{N}_0,
\end{equation}
for $m$ randomly chosen sites, with $m\ll n$. For $n=100$ we
consider $m=1$. For future reference we note that in numerical
simulations the important quantity is actually the ratio
$\bar{N}_c /\bar{N}_0$

\begin{equation}
\frac{\bar{N} _c}{\bar{N}_0}=6.67\times x^{-1}10^4\label{eq:relX}.
\end{equation}

To summarize, for a QCD axion with $\mu R_g = 0.42$, initial cloud population with $x=95\%$ and after
rescaling the values of $\bar{N} _c$ and $\bar{N} _0$
so that $\bar{N} _0$ is of order unity (for practical reasons) we
get the following parameters used in the simulation

\begin{eqnarray}
\bar{N} _c = 280000, \mbox{   }\bar{N} _0 = 4 \nonumber\\
n=100, \mbox{   }m=1.
\end{eqnarray}

In words, at each time step $k$, a number $100$ units are added to
the cloud and $\bar{N}$ grows locally. The local enhancement of
axion number is done randomly, i.e., $100$ randomly chosen grids
will receive one unit. At the same timestep, $1$ unit dissipates
from a randomly chosen site and it is assumed that this unit
escapes to infinity (it is not fed to the BH). Adding new axions
at \emph{randomly} chosen positions is a nontrivial assumption,
but it seems well justified based on two considerations: the
observations of laboratory Bosenovas which emphasize the
stochastic nature of this process~\cite{Gerton:2000} and the
previous statement that, until the critical limit is reached, the
system evolves in a linear regime, Eq.~\eqref{eq:linDiss}.

At each timestep $k$ all the grids are verified to see whether the
critical density $\bar{N} _c$ is reached. Once the criticality
condition is met, the gravitational potential is no longer
important and transfer of information is due to the nonlinear
interaction of the bosons. The next question is ``how many
neighbors does the site interact with?'' so that information of
collapse could be transported to that number of neighbors. We will
assume it interacts with 3 sites, namely those three which are
nearest neighbors and closer to the BH than the current site. The
critical condition and the critical flow are implemented as

\begin{eqnarray}\label{eq:crit1}
\mbox{if  }\bar{N} _{i,j}(k) > \bar{N} _c, \mbox{then  }\\
\nonumber \bar{N} _{i,j}(k)\to \bar{N} _{i,j}(k)-3\bar{N} _0;\\
\nonumber \bar{N} _{i+1,j}(k) \to \bar{N} _{i+1,j}(k)+\bar{N} _0;\\
\nonumber
\bar{N} _{i+1,j-1}(k) \to \bar{N} _{i+1,j-1}(k) +\bar{N} _0 ;\\
\nonumber
\bar{N} _{i+1,j+1}(k) \to \bar{N} _{i+1,j+1}(k) +\bar{N} _0 ;\\
\nonumber
\end{eqnarray}

\begin{figure}[!h]
\centerline{\includegraphics[width=\linewidth]{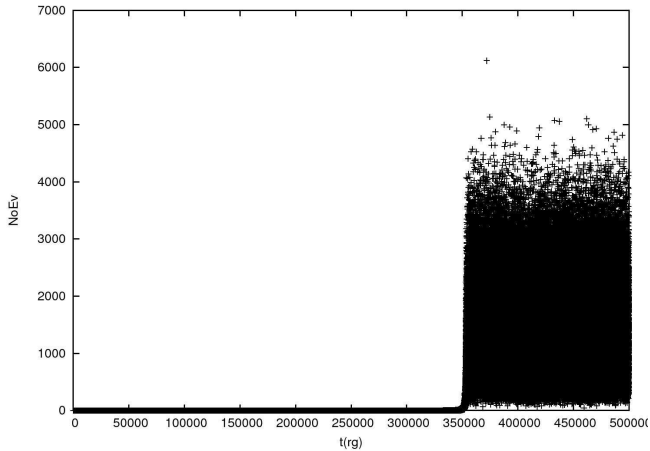}} \caption{The
distribution of event size, $D(s)$ for a QCD-axion cloud in
superradiance around a BH. The flow at criticality is given by the
rules~\eqref{eq:crit1}. The cloud was initialized at $95\%$ of its
critical boson number $ \bar{N} _c$.}\label{fig:soc1}
\end{figure}

The result of this simulation is the distribution $D(s)$ of event
sizes (Fig.~\ref{fig:soc1}). This is exactly how an archetypical
SOC system behaves~\cite{Bak:1988zz} and it is clear that after a growth
period \emph{Self Organized Criticality is reached and the system
stays in this state}. This is a first important result.

\begin{figure}[!h]
\centerline{\includegraphics[width=\linewidth]{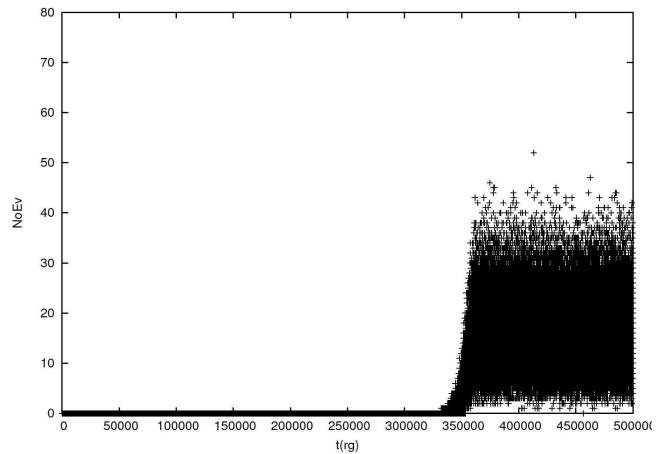}} \caption{The
distribution of event size, $D(s)$ for a QCD-axion cloud in
superradiance around a BH. The flow at criticality is given by
rules~\eqref{eq:crit2}. The cloud was initialized at $95\%$ of its
critical boson number.}\label{fig:soc2}
\end{figure}

To test whether or not the fact that the system reaches SOC is
rule-dependent, the system evolution is studied for different
rules at criticality (different critical flows). The difference is
that apart from the three units allowed to move to other sites, an
additional number of boson units completely leave the cloud. This
was done for $3$ units, $6$ units [Eq.~\eqref{eq:crit2},
Fig.~\ref{fig:soc2}] and $12$ units. For the case of six units,
this is written and implemented as

\begin{eqnarray}\label{eq:crit2}
\mbox{if  }\bar{N} _{i,j}(k) > \bar{N} _c, \mbox{then  }\\ \nonumber
\bar{N} _{i,j}(k)\to\bar{N} _{i,j}(k)-9\bar{N} _0;\\ \nonumber
\bar{N} _{i+1,j}(k) \to \bar{N} _{i+1,j}(k)+\bar{N} _0;\\ \nonumber
\bar{N} _{i+1,j-1}(k) \to \bar{N} _{i+1,j-1}(k) +\bar{N} _0 ;\\ \nonumber
\bar{N} _{i+1,j+1}(k) \to \bar{N} _{i+1,j+1}(k) +\bar{N} _0 ;\\ \nonumber
\end{eqnarray}

In all cases SOC is reached in a time period that depends on how
close $N_i$ is to the critical value. The number of events needed
to relax the system changes.

A more realistic simulation has to account for when the
superradiance condition is no longer fulfilled. It is then we
expect that SOC to cease, followed by a transitory period, and
afterwards the system will again settle into a SOC state
characterized by the new parameters. Due to the limited
computational capabilities and the very slow growth rate, we use
the fact that the system will eventually be in SOC and initialize
the automaton very close to its critical parameter, $x=0.99$.

To compute how much mass the BH has lost because of superradiance,
we assume as an example that superradiance stops when $10\%$ of
the BH mass is lost, i.e. superradiance stops for $M_{BH}=M$ but
immediately starts again for $M_{BH}=0.9M$. All the parameters
characteristic to the system, and thus to the simulation, change
accordingly. Simulations including the change in superradiance
parameters were performed for all 4 sets of rules at criticality.
The same qualitative results were obtained in all simulations:
there is SOC on both sides of the transition, the transition is
sharp and the ratio of the number of events prior to transition
and following transition is approximately $2$.

\begin{figure}
  \begin{center}
    \begin{tabular}{cc}
    {\includegraphics[width=\linewidth]{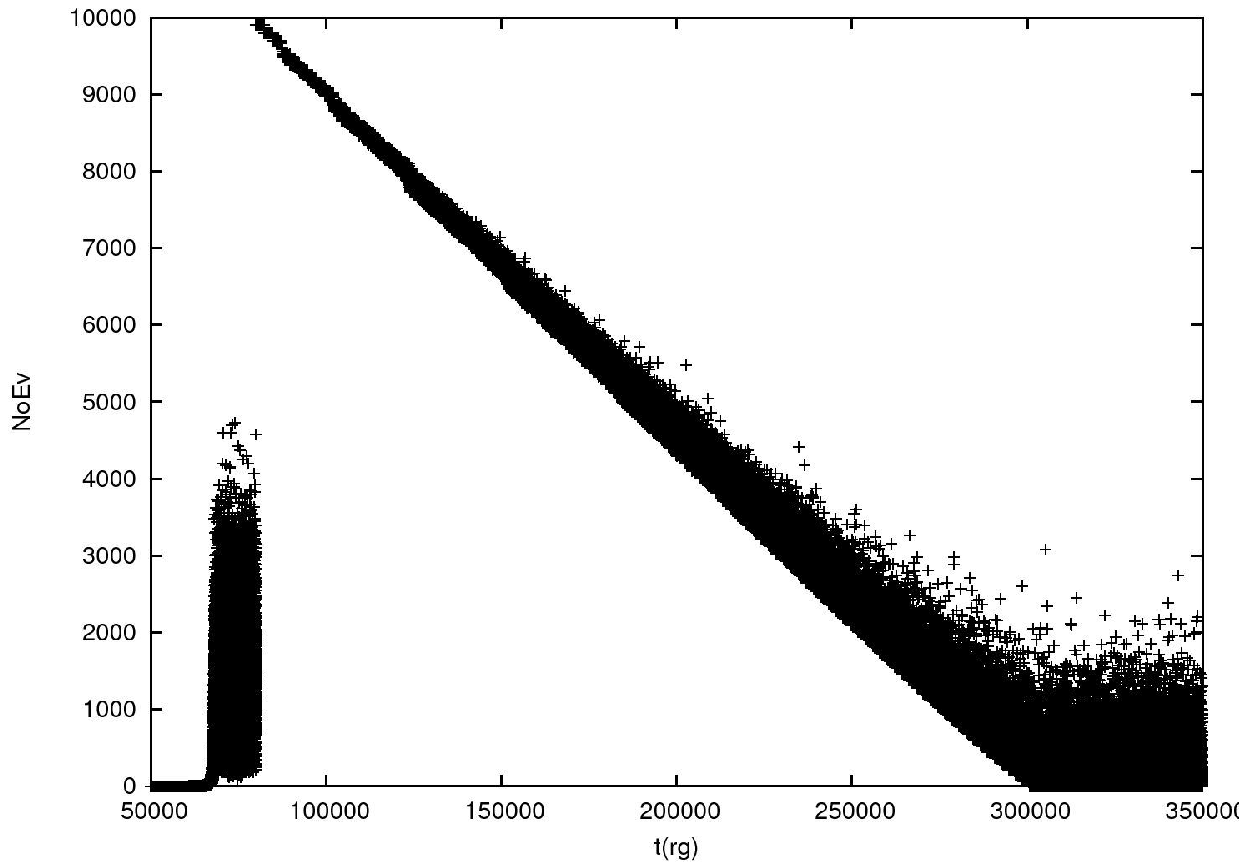}} \\
        {\includegraphics[width=\linewidth]{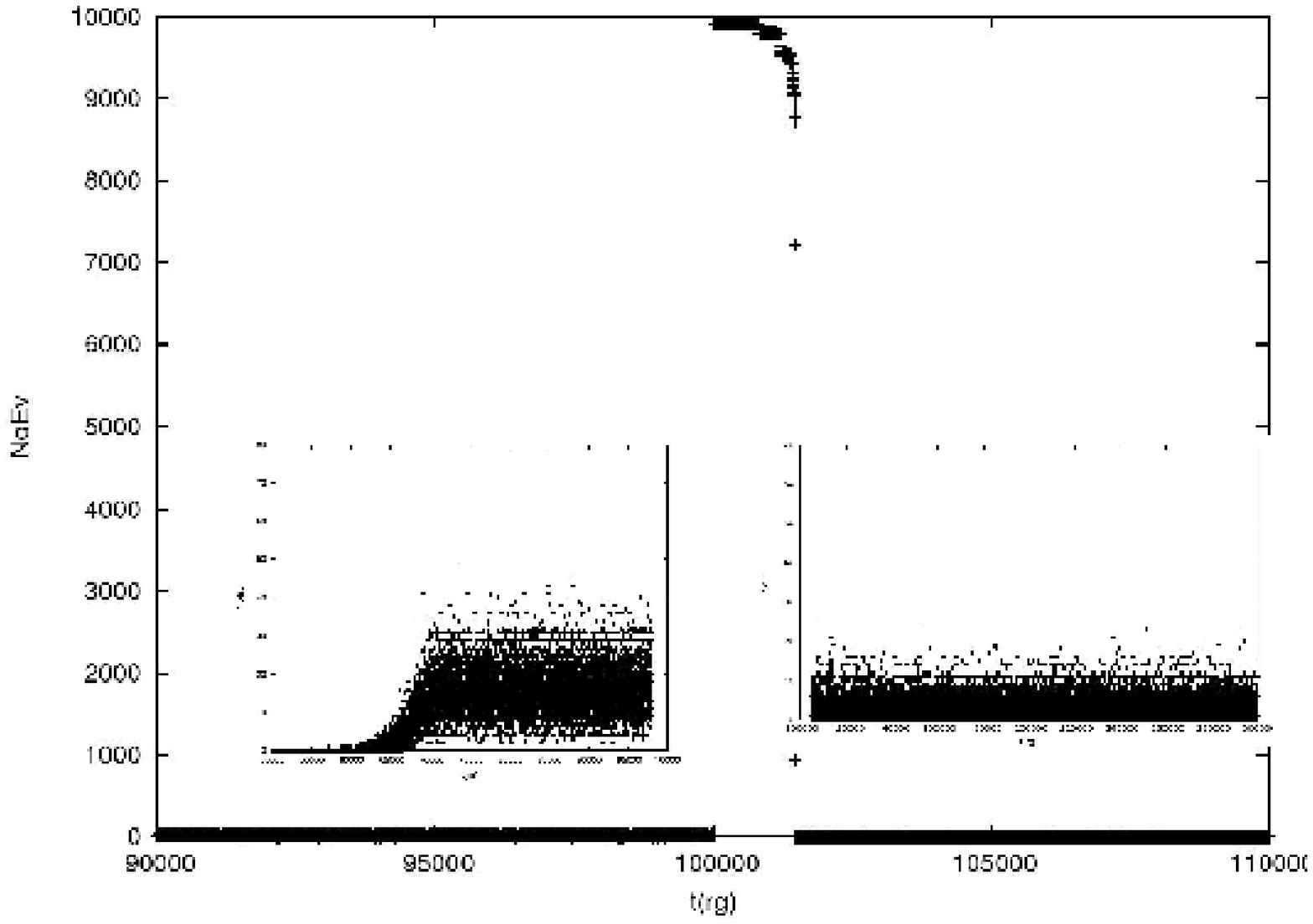}} \\
\end{tabular}
\caption{The distribution of event size, $D(s)$ for a QCD-axion
cloud in superradiance around a BH. The cloud was initialized at
$99\%$ of its critical boson number (the constraint
Eq.~\eqref{eq:relX} is obeyed). Transition to another set of
superradiance parameters is visible. The flow at criticality is
given by rules~\eqref{eq:crit1} (upper panel) and for
rules~\eqref{eq:crit2} (lower panel). }\label{fig:soc3}
  \end{center}
\end{figure}

When exiting the superradiance condition for the parameter set
governed by $M$, superradiance for the parameter set governed by
$=0.9M$ instantly begins according to our rules. The results of
this is that $\bar{N}$, which has a value around $\bar{N}_c^{M}$
is compared to the lower value of $\bar{N}_c^{0.9M}$ and
avalanches occur until the new value for the critical parameter is
reached and SOC is installed and maintained.

\section{Summary and open issues}\label{sec:5}


Starting from rigorous analytical results describing the dynamics
of a boson-cloud--black-hole (BC-BH) system, we derived a simple
map from this dynamics to a cellular automaton. Within our working
framework (see assumptions at the end of Section~\ref{sec:3} and
Table~\ref{table:map}) we have shown that the BC-BH system reaches
self-organized criticality (SOC) and stays there as long as the
superradiance condition is fulfilled. Since laboratory
observations of the growth and collapse of Bose--Einstein
condensates lead to a similar picture
--- albeit only for short timescales, see for instance Fig.~3 in \cite{Gerton:2000} --- we have some
confidence that this is a physical property of our model and not
an artifact of inadequate modelling. When the superradiance
condition is no longer valid, the evolution rules change
quantitatively and the simulated system exhibits a transition
after which it again settles into SOC. By varying the rules we
checked that this is a rule-\emph{independent} result (in the
sense that it is recovered for all the sets of rules that we have
employed) which gives some confidence that this result holds for
actual axion cloud near black holes.


The quantitative details of the transition are rule dependent
(Figs.~\ref{fig:soc1}-\ref{fig:soc3}). For the first set of
transition rules, Eq.~\eqref{eq:crit1}, the temporal width of the
transition period is quite large, $25\times 10^4r_g$. For the
second set of transition rules, Eq.~\eqref{eq:crit2}, the temporal
width of the transition is of the order $10^4r_g$. The ratio of
the medium numerical value of $D(s)$ before and after transition
is approximately $2$ for four different evolutions of the system
at criticality. The open issue is whether or not these aspects are
related to underlying physics or are numerical artifacts.

We assumed that a new superradiance domain is instantly valid,
which led to very large number of events associated with the
transition  when compared to the SOC state (Fig.~\ref{fig:soc3}).
An open question is whether or not this assumption is physically
justified. This should be investigated in future work.

We expect that observational data of gravitational waves from a
large range of masses for BH will be able to shed light on at
least two very important issues. One is the physics at transition
and observational data will help devise more realistic rules for
suitable cellular automata. Also, an appropriate observational
database will help in determining the scaling function associated
with this phenomena.

\acknowledgments

We are grateful to Asimina Arvanitaki, Sergei Dubovsky and Niklas
Johansson for helpful comments. DG thanks Frank Wilczek for
drawing his attention to axion clouds around black holes during
the conference ``Strings 2011'' in Uppsala. GM thanks \c{S}tefan
Petrea for valuable help in optimizing the code used to produce
the results in the simulation section.

GM acknowledges the financial support of the Sectoral Operational
Programme for Human Resources Development 2007-2013, co-financed
by the European Social Fund, under the project number
POSDRU/107/1.5/S/76841 with the title "Modern Doctoral Studies:
Internationalization and Interdisciplinarity". DG is supported by
the START project Y435-N16 of the Austrian Science Fund (FWF).

\begin{appendix}

\section{Sandpile model}

\begin{figure}[!h]
\centerline{\includegraphics[width=\linewidth]{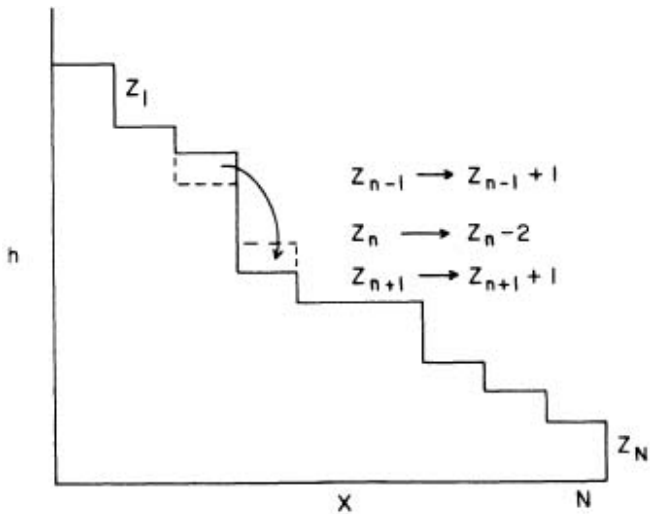}} \caption{One
dimensional sand pile automaton~\cite{Bak:1988zz}.}\label{fig:1DCA}
\end{figure}

SOC may be studied through a simple sand-pile model
(Fig.~\ref{fig:1DCA}) in a CA framework~\cite{Bak:1988zz}. This
arrangement may be thought of as half of a symmetric sand pile
with both ends open. The numbers $z_n$ characterizing the
automaton represent height differences between successive
positions along the sand pile, $z_n = h(n)-h(n+1)$. When placing a
grain of sand in cell $n$ the dynamics of this system follows the
rules

\begin{eqnarray}
z_n && \to z_n + 1\label{eq:AddGrain} \\  z_{n-1} && \to z_{n-1}
-1\nonumber.
\end{eqnarray}

This model is a cellular automaton where the state of the discrete
variable $z_n$ at time $t+1$ depends on the state of the variable
and its neighbors at time $t$.

When the height difference becomes higher than a threshold
(critical) value, $z_c$, one unit of sand tumbles to the lower
level

\begin{eqnarray}
z_n &&\to z_n - 2\label{eq:crit} \\  z_{n\pm1}&& \to
z_{n\pm1}+1\text{ for  }z_n>z_c.\nonumber
\end{eqnarray}

The process continues until all the $z_n$ are below $z_c$.  At
this point another grain of sand is added at a random site $n$
through Eq.~\eqref{eq:AddGrain}.

Starting from a set of initial
conditions, the pile grows and in the mean time the slope of the
pile increases. It is the value of this slope, as seen in the
Eqs.~\eqref{eq:AddGrain}, \eqref{eq:crit}, that will at some
point reach a critical value. The meaning of the critical value is
that if this value is reached and one more grain is added,
material will slide. In the critical state the system is just
barely stable with respect to further perturbations. Information
about the perturbation will be known across all lengthscales in
the system.

An \emph{event} consists of the seed (i.e. placing of
the grain) and the subsequent information (grain) transfer or
avalanche until complete relaxation. Each event has an associated
time $T$ (i.e. the time it takes the perturbation to die out), a
cluster size $s$ (i.e. the total number of slidings induced by the
perturbation) and a total energy (i.e. release of energy, as
transfers occur based on first principles that the final state is
an energetically lower state).

For this simple model, the distribution of event sizes $D(s)$ is given by~\cite{Pruessner}
\begin{equation}
D(s) = \frac{1}{L} \theta \left ( L-s \right ),
\end{equation}
where $L$ is the system size and $\theta (x)$ is the Heaviside step-function. To put this in the form of Eq.~\eqref{eq:distrib}, we rewrite it as

\begin{equation}
D(s) = \frac{1}{s} \left[ \frac{s}{L} \theta \left ( 1 - \frac{s}{L} \right ) \right ]
\end{equation}
and by identification $a=1$, $\tau = 1$, $b=1$, $\sigma = 1$ and the scaling function is $\mathcal{G}(x) = x \theta \left ( 1-x\right )$.

\end{appendix}

\bibliography{review}

\end{document}